# Dynamical-parameter algorithm for U(1) gauge theory [*]

Werner Kerler[a], Claudio Rebbi[b] and Andreas Weber[a]

[a] *Fachbereich Physik, Universität Marburg, D-35032 Marburg, Germany*
[b] *Department of Physics, Boston University, Boston, MA 02215, USA*

## Abstract

We present an algorithm for Monte Carlo simulations which is able to overcome the suppression of transitions between the phases in compact U(1) lattice gauge theory in 4 dimensions.

## 1. Introduction

Monte Carlo simulations of compact U(1) lattice gauge theory in 4 dimensions near its phase transition are hampered by the strong suppression of transitions between the phases. On larger lattices conventional algorithms are not able to induce transitions at all.

We obtain a more appropriate algorithm by using the Wilson action supplemented by a monopole term [1]

$$S = \beta \sum_{\mu > \nu, x} (1 - \cos \Theta_{\mu\nu,x}) + \lambda \sum_{\rho, x} |M_{\rho,x}|$$

[*] Contribution to LATTICE 94, International Symposium on Lattice Field Theory, Bielefeld, Germany, 1994. Research supported in part under DFG grants Ke 250/7-2 and 250/11-1 and under DOE grant DE-FG02-91ER40676

where $M_{\rho,x}$ is the monopole content of 3D cubes [2]. The strength of the first order transition decreases with $\lambda$, the transition ultimately getting of second order [3]. Therefore, by making $\lambda$ a dynamical variable, we no longer rely on the tunneling between the phases but get a much easier pathway along the mountains of the joint probability distribution $P(E, \lambda)$, where $E$ denotes the plaquette energy (Figure 2 gives an example of $P(E, \lambda)$). Our algorithm has the further virtue to be vectorizable and parallelizable.

Before running the dynamical-parameter algorithm one has to determine the phase transition line (shown in Figure 1) and the additional term in the action related to the prescribed distribution of $\lambda$. It is crucial that this can be done with low computational cost also on larger lattices. Here we present a method by which this is achieved.

## 2. Outline of method

Conventional methods simulate the probability distribution

$$\mu_\lambda(\Theta) = \exp(-S_\lambda(\Theta))/Z_\lambda$$

where $\lambda$ is a fixed parameter. In order to make $\lambda$ a dynamical variable we consider $\mu_\lambda(\Theta)$ as the conditioned probability to get a configuration $\Theta$ given a value of $\lambda$ and prescribe a probability distribution $f(\lambda)$ to get the joint probability distribution $\mu(\Theta, \lambda) = f(\lambda)\mu_\lambda(\Theta)$. To simulate $\mu(\Theta, \lambda)$ we need it in the form

$$\mu(\Theta, \lambda) = \exp(-S(\Theta, \lambda))/Z \qquad (1)$$

where $S(\Theta, \lambda) = S_\lambda(\Theta) + g(\lambda)$ and $g(\lambda)$ is determined by $f(\lambda)$.

In our application of the algorithm each update of the link variables $\Theta_{\mu,x}$ is followed by an update of $\lambda$, which is done in a discrete set of values $\lambda_q$ with $q = 1, \ldots, n$. The individual update steps are Metropolis steps in both cases. For the $\lambda$ update the proposal matrix $\frac{1}{2}(\delta_{q+1,q'} + \delta_{q,q'+1} + \delta_{q,1}\delta_{q',1} + \delta_{q,n}\delta_{q',n})$ and the acceptance probabilities $\min(1, \exp(S(\Theta, \lambda_q) - S(\Theta, \lambda_{q'})))$ are used.

For each value of $q$ one needs $\beta(\lambda_q)$ and $g(\lambda_q)$. We require $\beta(\lambda_q) \approx \beta_w(\lambda_q)$ where $\beta_w$ is the $\beta$ value where both phases are equally probable. Our condition for fixing $g(\lambda_q)$ is $f(\lambda) \approx$ const. In order to determine the sets of $\beta(\lambda_q)$ and $g(\lambda_q)$ we use the fact that in a simulation the transition probabilities between neighboring values of $\lambda$ are very sensitive to these quantities.

## 3. Transition probabilities

To derive relations which can be used in the envisaged determination of $\beta(\lambda_q)$ and $g(\lambda_q)$ we use the probability for the transition from a value $\lambda_q$ to a neighboring value $\lambda_{q'}$

$$W(\Theta, q; q') = \frac{1}{2}\min(1, \exp(S(\Theta, \lambda_q) - S(\Theta, \lambda_{q'}))) \qquad (2)$$

and note that detailed balance implies

$$f(\lambda_{q-1})\mu_{\lambda_{q-1}}(\Theta)W(\Theta, q-1; q) = f(\lambda_q)\mu_{\lambda_q}(\Theta)W(\Theta, q; q-1) . \qquad (3)$$

Introducing the average transition probability for a set $K(q)$ of $\Theta$ configurations

$$p_K(q; q') = \frac{1}{w_K(q)} \sum_{\Theta \in K} \mu_{\lambda_q}(\Theta)W(\Theta, q; q') , \qquad (4)$$

where $w_K$ is the weight of this set, we get

$$f(\lambda_{q-1})\ w_K(q-1)\ p_K(q-1; q) = f(\lambda_q)\ w_K(q)\ p_K(q; q-1) . \qquad (5)$$

from averaging (3).

We now apply (5) to sets of configurations in the hot phase and in the cold phase separately. Because we are interested in cases where transitions between the phases are extremely rare, sets of this type with numbers of configurations sufficient for the present purpose are easily obtained. The respective equations can be considered to be independent. Using (5) for $K = K_c$ and $K = K_h$ we get a pair of equations which simplifies to

$$\begin{aligned} p_{K_c}(q-1; q) &= p_{K_c}(q; q-1) \\ p_{K_h}(q-1; q) &= p_{K_h}(q; q-1) \end{aligned} \qquad (6)$$

if $\beta = \beta_w$ and $f(\lambda)$ =const. This is what we exploit to determine $\beta(\lambda_q)$ and $g(\lambda_q)$.

Our strategy is to adjust $\beta(\lambda_q)$ and $g(\lambda_q)$ for known $\beta(\lambda_{q-1})$ and $g(\lambda_{q-1})$ such that (6) holds. Starting from given $\beta(\lambda_1)$ and arbitrarily chosen $g(\lambda_1)$ in this way we can obtain $\beta(\lambda_q)$ and $g(\lambda_q)$ for $q = 2, \ldots, n$.

## 4. Determination of $\beta(\lambda)$ and $g(\lambda)$

To start our procedure we use a value of $\lambda_1$ in the region where the peaks of the probability distribution related to the phases strongly overlap so that tunneling is no problem and $\beta(\lambda_1)$ can easily be obtained by a conventional simulation. Because only the differences $g(\lambda_{q-1}) - g(\lambda_q)$ are relevant we can choose $g(\lambda_1) = 0$. Then for $q = 2, \ldots, n$ we consecutively determine $\beta(\lambda_q)$ and $g(\lambda_q)$ for known $\beta(\lambda_{q-1})$ and $g(\lambda_{q-1})$.

Within a step from $q-1$ to $q$ we first get a rough approximation of $\beta(\lambda_q)$ by extrapolation from former values and an approximate new $\lambda_q$ in about the same distance as in former steps. Then we generate the sets of $\Theta$ configurations $K_c$ and $K_h$ by short Monte Carlo runs with cold and hot start, respectively. Because for a set $K(q)$ the quantity

$$\tilde{p}_K(q; q') = \frac{1}{N_K(q)} \sum_{\Theta \in K} W(\Theta, q; q'),  \tag{7}$$

where $N_K$ is the number of configurations, approximates (4), this allows us to calculate approximate quantities which should satisfy (6). These are two equations determining the two unknown values $\beta(\lambda_q)$ and $g(\lambda_q)$. We obtain good estimates for them though only approximate quantities enter (6) because the peaks related to the phases vary only little with $\beta$. In addition, $\tilde{p}(q-1; q)$ is used to adjust the distances between neighboring $\lambda$ values such that one has roughly equal transition probabilities for all steps.

After a larger number of $q$ steps the errors may accumulate. Therefore, we perform short runs of the dynamical-parameter algorithm to test wether it does indeed travel along the mountains of the distribution in the hot as well as in the cold phase. If it gets stuck we slightly increase or decrease the $\beta(\lambda)$ in the region of $\lambda$ where the transitions fail. We then determine the corresponding values of $g(\lambda_q)$ from the conditions

$$\tilde{p}_{K_c}(q-1; q) + \tilde{p}_{K_h}(q-1; q) = \tilde{p}_{K_c}(q; q-1) + \tilde{p}_{K_h}(q; q-1)  \tag{8}$$

and try again to run the dynamical algorithm. Typically one or two trials are sufficient.

After performing the actual simulations with dynamical $\lambda$, improved $\beta(\lambda_q)$ can be obtained by reweighting [5] the distribution at $\lambda$ values where deviations from equal weights of the phases occur. The related new $g(\lambda_q)$ then are obtained from (8). The values $g(\lambda_q)$ can be improved by replacing them by $g(\lambda_q) + \ln(f(\lambda_q))$.

## 5. Numerical results

Figure 1 shows the location of the maximum of the specific heat $\beta_C$ as function of $\lambda$. The method presented has enabled us to determine this phase transition line also for lattice size $16^4$ needing only a small amount of computer time (which by conventional simulations due to the suppression of transitions is not possible at all). To get the sets of configurations $K_c$ and $K_h$ we used about $2 \times 10^4$ sweeps per $\lambda$ and in the short test runs with the dynamical algorithm $4 \times 10^4$ sweeps in total. The calculations have been done for $n = 25$ values of $\lambda_q$ ranging from $\lambda = 0$ to $\lambda = 0.4$.

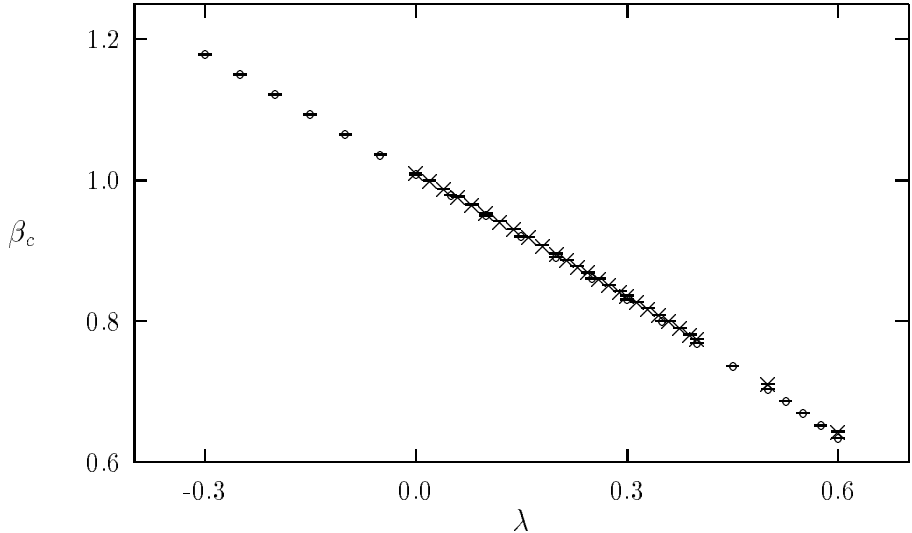

Figure 1: Location of phase transition as function of $\lambda$ for $8^4$ (circles) and $16^4$ (crosses) lattices.

In our simulations with dynamical $\lambda$, in which we collected about $10^6$ sweeps, it has become possible to have transitions between the phases also on the $16^4$ lattice. They confirmed the values which we have found for $\beta_w(\lambda_q)$ by the procedure described above. Figure 2 shows the (reweighted) distribution $P(E, \lambda)$ which we obtained in our simulations (by the dynamical-parameter algorithm except for $\lambda = 0.5$ and $\lambda = 0.6$).

On the $8^4$ lattice there is still substantial tunneling. This is reflected by the overlap of the peaks in the distribution $P(E, \lambda)$ which we have presented in [3]. There the $g(\lambda_q)$ have been determined by a simpler method based on (8).

The tunneling times between the phases in our algorithm with dynamical $\lambda$ on the $8^4$ lattice are greatly reduced as compared to those of a conventional Metropolis algorithm [3]. On the $16^4$ lattice, where for the conventional algorithm one observes no transitions at all, for $\lambda = 0$ we get a time of the order of $10^3$ for our algorithm.

For further reduction of the autocorrelation times in addition to avoiding tunneling one would have to replace the local Metropolis steps for $\Theta$ by more efficient ones

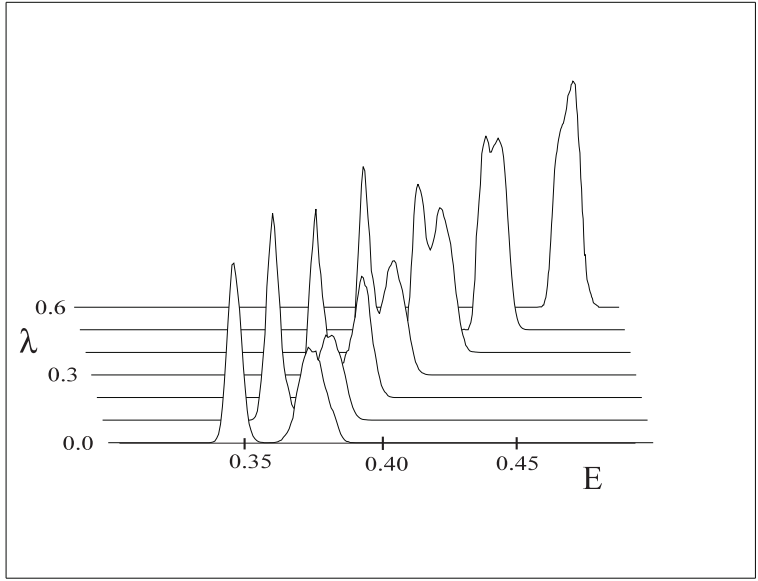

Figure 2: Distribution $P(E,\lambda)$ at the phase transition line on $16^4$ lattice.

(corresponding to the cluster steps in the analogous algorithm for the Potts model [6]).